%% file: IASD_jrnl.tex
\documentclass[peerreview,draftcls,onecolumn,12pt]{IEEEtran}
\usepackage{amsmath,amssymb,latexsym,epsfig,epic}
\usepackage[usenames]{color}
\usepackage{subfigure}
\usepackage{theorem}
\usepackage{relsize}
\usepackage{cite}
\usepackage{booktabs}

\usepackage{graphicx}
\usepackage{epsf}
\usepackage{latexsym}
\usepackage{amsmath}
\usepackage[normalem]{ulem}
\usepackage{algorithm}
\usepackage{algorithmic}
\usepackage{verbatim}

\newcommand{\abs}[1]{\left\lvert#1\right\rvert}
\newcommand{\norm}[1]{\left\lVert#1\right\rVert}
\newcommand{\bracket}[1]{\left(#1\right)}

\newcommand{\figref}[1]{Fig.~\ref{#1}}

\newcommand{\secref}[1]{Sec.~\ref{#1}}

\begin{document}

\title{Interference Mitigation via \\Interference-Aware Successive Decoding
\footnote{The material in this paper was presented in part to 2012 Information Theory and Applications Workshop, Feb. 2012}}

\author{Hyukjoon~Kwon,~\IEEEmembership{Member,~IEEE},
Jungwon~Lee,~\IEEEmembership{Senior Member,~IEEE},
and Inyup~Kang,~\IEEEmembership{Member,~IEEE}
\thanks{H.~Kwon, J.~Lee and I.~Kang are with Mobile Solution Lab, Samsung US R\&D Center, San Diego, CA, 92130, U.S.A. (e-mail: \{hyukjoon, jungwon\}@alumni.stanford.edu, inyup.kang@samsung.com)}}

\markboth{Submission to IEEE Transactions on Wireless Communications}{Shell \MakeLowercase{\textit{et al.}}: Bare Demo of IEEEtran.cls for Journals}
\maketitle

\begin{abstract}
In modern wireless networks, interference is no longer negligible since each cell becomes smaller to support high throughput. The reduced size of each cell forces to install many cells, and consequently causes to increase inter-cell interference at many cell edge areas. This paper considers a practical way of mitigating interference at the receiver equipped with multiple antennas in interference channels. Recently, it is shown that the capacity region of interference channels over point-to-point codes could be established with a combination of two schemes: treating interference as noise and jointly decoding both desired and interference signals. In practice, the first scheme is straightforwardly implementable, but the second scheme needs impractically huge computational burden at the receiver. Within a practical range of complexity, this paper proposes the interference-aware successive decoding (IASD) algorithm which successively decodes desired and interference signals while updating \emph{a priori} information of both signals. When multiple decoders are allowed to be used, the proposed IASD can be extended to interference-aware parallel decoding (IAPD). The proposed algorithm is analyzed with extrinsic information transfer (EXIT) chart so as to show that the interference decoding is advantageous to improve the performance. Simulation results demonstrate that the proposed algorithm significantly outperforms interference non-decoding algorithms.
\end{abstract}



\section{Introduction} \label{sec:intro}
\input{intro.tex}

\section{System Model} \label{sec:model}
\input{system.tex}

\section{Reviews of Interference Mitigation Algorithms} \label{sec:algorithm}
\input{reviews.tex}

\section{Interference-Aware Decoding} \label{sec:IAD}
\input{IADec.tex}

\section{Extrinsic Information Transfer Chart} \label{sec:analysis}
\input{complexity.tex}

\section{Simulation Results} \label{sec:result}
\input{results.tex}

\section{Conclusion} \label{sec:conclusion}
\input{conclusion.tex}



\bibliographystyle{IEEEtran}
\bibliography{IASD_jrnl}

\end{document}

%% file: intro.tex

The recent wireless environment has rapidly changed to support high throughput as well as reliable communication. This advancement of wireless channels leads to reduce the cell size in cellular networks or makes local cells such as pico/femto-cells favorable under a macro-cell. When the cell size is reduced, it is advantageous that each base station (BS) can transmit signals with high order modulation and high transmit power. On the other hand, it is disadvantageous that interference is no longer negligible at many cell-edge areas. This causes inter-cell interference to be more important for determining the average throughput in a cell.

In order to overcome the adverse effect of inter-cell interference, the direct channel from a BS to the served mobile station (MS) should be fundamentally robust. A variety of techniques have been researched to achieve the capacity in a point-to-point channel. \cite{Telatar99} reveals that the capacity of multi-antenna channels significantly increases over the benefit of spatial diversity gain. Such a spatial gain has been achieved from many algorithms. For instance, \cite{Alamouti98} proposes to transform channels to be orthogonal over two time slots. \cite{Tarokh99} uses space-time block codes to maximize the diversity order. Moreover, \cite{Foschini96} designs a transceiver to exploit multiple paths for sending data streams between the transmitter and the receiver, which is called Bell Lab layered space-time (BLAST). It is extended to the vertical BLAST (V-BLAST) detecting algorithm in \cite{Wolniansky98} to eliminate inter-stream interference successively.

The space-time detection schemes has been researched in conjunction with channel coding schemes, and evolve to iterative algorithms called iterative detection and decoding (IDD). Under the IDD-structured receiver, both the detector and the decoder are connected with feedforward and feedback channels, and exchange extrinsic information that is expressed as the logarithm of the ratio of bit probabilities called \emph{L}-value or log-likelihood ratio (LLR) \cite{Hagenauer96}. The IDD techniques are studied with many detecting algorithms to mitigate inter-stream interference. \cite{Tonello00} directly maximizes \emph{a posteriori} (MAP) LLRs from bit-interleaved coded modulation (BICM) systems. Even though the MAP detector is shown to be optimal, its computation complexity is extremely huge with multiple antennas or high order modulations. In order to alleviate the computational burden, the sphere decoding technique is designed on the principle to take into account only a list of candidate symbols \cite{Fincke85,Hochwald03}. In addition, a linear minimum mean square error (MMSE) detector has been analyzed, which is applicable for less-computation receivers under the IDD structure \cite{Tuchler02MMSE,junwon10}.

However, such a simple method only reinforcing the performance of point-to-point channels is limited to improve the performance of interference channels. For the next generation wireless communication, it is necessary to mitigate interference signals at the MS or efficiently manage the interference at the BS. The most common method used for interference mitigation is to apply orthogonal multiplexing schemes to the system. For example, a cellular network has an ability to re-use frequencies among cells placed far away with the adjacent cells using different frequencies \cite{Wang02}. As adjusting the frequency reuse factor defined as the rate at which the same frequency is used, the cellular network can mitigate the interference so as to increase both coverage and capacity. Another orthogonal multiplexing scheme is a code division multiple access (CDMA) where multiple MSs are allowed to be multiplexed over the same channel \cite{Gilhousen91}. Using the CDMA technique, each BS assigns its own code so that the signal from the BS can be spread out with the unique code. Even though these orthogonal multiplexing schemes are effective to avoid interference signals, they have a drawback that they cannot achieve the full degrees of freedom available in the channel. Thus, there have been many efforts for achieving universal frequency reuse without using orthogonal multiplexing schemes.

The universal frequency reuse enables cellular networks to be modeled as interference channels that have been studied extensively in information theory. Most of theoretical researches have been mainly focused on finding the outer bound of the capacity region in interference channels, the transmission scheme applicable for reaching the outer bound, or the encoding scheme to achieve near-capacity. Relatively less emphasis has been placed on the analysis of interference channels when point-to-point codes are used with the advanced receiver. Recently, there have been active researches in the direction that advanced receivers mitigate interference \cite{WAYV07, BJ09, BET11}. In particular, \cite{BET11} shows that the capacity region of an interference channel with point-to-point codes can be established by a combination of treating interference as noise and using joint decoding of the desired signal with interference.

This paper examines how the advanced receiver should be designed in order to effectively mitigate interference signals when each MS is constrained to use a point-to-point code in multi-input multi-output (MIMO) interference channels. Simply, interference can be treated as noise \cite{Winters84}. Then, aforementioned techniques designed for a point-to-point channel can be directly applicable. Alternatively, both desired and interference signals can be jointly decoded. However, it is extremely complicated so that the joint decoding technique is impractical for a MS in modern cellular networks. Instead, this proposes another interference decoding algorithm named interference-aware successive decoding (IASD). The proposed IASD operates with a reasonable range of complexity while achieving the performance improvement over conventional schemes. Along with IASD, this paper also proposes an interference-aware parallel decoding (IAPD) algorithm that improves the performance more at the cost of multiple decoders and high complexity. Both IASD and IAPD will be described in detail. 

This paper is organized as follows: Sec.~\ref{sec:model} describes the system model in an interference channel, and Sec.~\ref{sec:algorithm} reviews the existing interference mitigation algorithms: interference-whitening and interference-aware detectors. In Sec.~\ref{sec:IAD}, the proposed interference decoding algorithms are explained with their IDD structure in the receiver. Sec.~\ref{sec:analysis} exploits the extrinsic information transfer chart to analyze the proposed algorithm with existing ones.  Sec.~\ref{sec:result} evaluates the performance of the proposed algorithms and compares with analytical results. The conclusion is followed in Sec.~\ref{sec:conclusion}.

%% file: system.tex

This paper considers a Gaussian interference channel (IC) where $U$ base stations (BSs) are transmitting their own messages to the designated mobile stations (MSs), respectively. The BS serving for the $u$th MS transmits $N_{s,u}$ spatial streams over $N_{t,u}$ antennas. The MS is equipped with $N_{r,u}$ antennas. The spatial multiplexing scheme is used over multiple antennas. For simplicity, the number of antennas for each BS and MS is assumed to be same, and the subscript $u$ is omitted in the following. In general, this channel model is considered as a multi-cell multi-user system so that inter-cell interference is not negligible at the cell edge. Each BS uses its own point-to-point code to maximize the performance for the designated MS. For this reason, this paper assumes that a bit interleaved coded modulation (BICM) scheme is used in conjunction with binary codes such as convolutional codes, turbo codes, and low-density parity check codes.

\begin{figure*}[htb]
\centering
\includegraphics[width=6in]{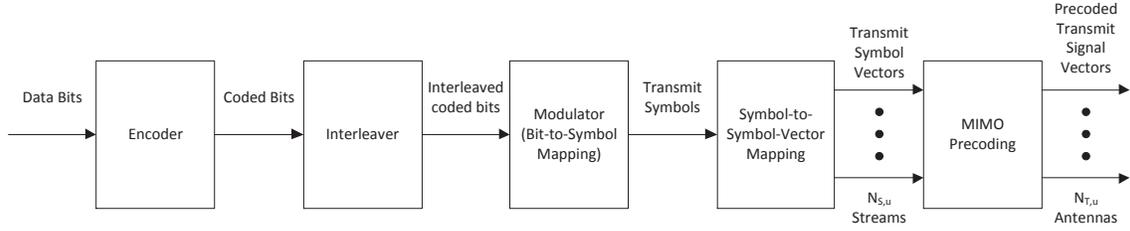}
\caption{A block diagram of transmitter using a point-to-point code over BICM}
\label{fig:transmitter}
\end{figure*}

\figref{fig:transmitter} illustrates the block diagram of a BS using spatial multiplexing with BICM. Initially, the BS encodes the sequence of binary data bits per codeword. The coded bits, lengthened by the code rate, are scrambled and permuted at the interleaver. A large interleaving pattern would generate statistically independent bits. The coded sequence is grouped into $N$ bits and mapped to $M$-ary quadrature amplitude modulation (QAM) symbols on the constellation. Then, a group of $N_s$ symbols form a transmit symbol vector that is precoded to form a transmit signal vector. Finally, this signal vector is transmitted over multiple antennas. The precoding matrix is combined with channel matrices such that only the combined equivalent channel is detected at the MS. No precoding can also be represented as a trivial linear precoding with an identity matrix.

At the $u$th MS, the received signal $\mathbf{y}_u$ is given by a $N_r \times 1$ vector as
\begin{eqnarray}
\mathbf{y}_u = \sqrt{P_u}\mathbf{H}_{uu}\mathbf{x}_u + \sum_{\substack{i=1,i \neq u}}^U \sqrt{P_i}\mathbf{H}_{ui}\mathbf{x}_i + \mathbf{n}_u
\end{eqnarray}
where $\mathbf{H}_{ij}$ is the channel matrix from the $j$th BS to the $i$th MS. The $s$th column $\mathbf{h}_{ij,s}$ of $\mathbf{H}_{ij}$ corresponds to the channel vector for the $s$th stream of the $i$th MS from the $j$th BS. The signal $\mathbf{x}_i$ is a transmit symbol vector for the $i$th MS and its power is normalized, i.e., $E\left[ \mathbf{x}_i\mathbf{x}_i^\dagger \right] = I_{N_s}$. $P_i$ denotes the effective channel gain combined with the transmit power from the $i$th BS. This paper assumes that the channel is quasi-static, consisting of independent fading blocks. The noise $\mathbf{n}_u$ is an independent and identical distributed (i.i.d.) complex Gaussian random vector with zero mean and identity covariance. Thus, given the channel matrix and the received signal, the conditional probability density function (PDF) of the noise vector $\mathbf{n}_u$ is expressed as
\begin{eqnarray}
f_{\mathbf{n}_u | \mathbf{H}_{ui}, \mathbf{y}_u \: \forall i} & \nonumber \\
&\hspace{-1cm}= \frac{1}{\pi^{N_r}}\exp\left( -\norm{ \mathbf{y}_u - \sum_{\substack{i=1}}^U \sqrt{P_i}\mathbf{H}_{ui}\mathbf{x}_i }^2 \right).
\end{eqnarray}
In the next sections, this paper investigates how to mitigate the interference at the $u$th MS under the knowledge of channel matrices $\mathbf{H}_{ui}$ for all $i$. These algorithms can be straightforwardly applied from a single interference to multiple interferences. For ease of analysis, $U = 2$ is considered such that each MS handles the desired signal with a single interference. Then, the received signal can be rewritten as
\begin{eqnarray}
\mathbf{y} = \sqrt{P_D}\mathbf{H}^{D}\mathbf{x}^D + \sqrt{P_I}\mathbf{H}^I\mathbf{x}^I + \mathbf{n} \label{eq:DIeq}
\end{eqnarray}
where the script $D$ and $I$ denote the desired and interference signals, respectively. The signal-to-noise ratio (SNR) and the signal-to-interference ratio (SIR) for each stream simply becomes $P_D$ and $P_D/P_I$, respectively.

%% file: reviews.tex

This section reviews two existing interference mitigation algorithms: one is simply treating interference as Gaussian noise, and the other is taking advantage of the fact that interference symbols are discretely arranged on the constellation. Both algorithms have been designed for improving the detection performance of the desired signal. In addition, both are non-iterative algorithms that require no \emph{a priori} information.

\subsection{Interference-Whitening (IW) Detection} \label{sec:IW}
\input{IW.tex}

\subsection{Interference-Aware Detection (IA-Detection)}  \label{sec:IADet}
\input{IADet.tex}

%% file: IW.tex

Among many algorithms to handle interference, the straightforward approach is to treat interference as Gaussian noise \cite{Winters84}. Then, the effective noise combined with interference is assumed to be a colored Gaussian noise. Under this approach, the receive signal in \eqref{eq:DIeq} can be represented as
\begin{eqnarray}
\mathbf{y} = \sqrt{P_D}\mathbf{H}^{D}\mathbf{x}^D + \mathbf{v} \label{eq:IWeq}
\end{eqnarray}
where $\mathbf{v}$ is the effective noise vector integrated with interference, i.e., $\mathbf{v} = \sqrt{P_I}\mathbf{H}^{I}\mathbf{x}^I + \mathbf{n}$. Since interference signals are modulated on the symmetric constellation, the expectation of $\mathbf{x}^I$ is zero so that $\mathbf{v}$ also has the zero mean. Accordingly, the covariance matrix of $\mathbf{v}$ is calculated as
\begin{eqnarray}
\mathbf{R}_{\mathbf{v}} = P_I \mathbf{H}^I\mathbf{H}^{I\dagger} + \mathbf{I}_{N_r}
\end{eqnarray}
such that \eqref{eq:IWeq} is whitened as multiplying $\mathbf{R}_{\mathbf{v}}^{-\frac{1}{2}}$ by the receive signal $\mathbf{y}$. Then, the whitened channel becomes
\begin{align}
\tilde{\mathbf{y}} &= \mathbf{R}_{\mathbf{v}}^{-\frac{1}{2}}\mathbf{y} \nonumber \\
&= \sqrt{P_D}\mathbf{R}_{\mathbf{v}}^{-\frac{1}{2}}\mathbf{H}^D \mathbf{x}^D + \tilde{\mathbf{v}} \nonumber \\
&= \sqrt{P_D}\tilde{\mathbf{H}}^D \mathbf{x}^D + \tilde{\mathbf{v}} \label{eq:IWptp}
\end{align}
where $\tilde{\mathbf{H}}^D$ is the effective desired channel matrix. Also, the whitened noise vector is defined as
\begin{eqnarray}
\tilde{\mathbf{v}} = \mathbf{R}_{\mathbf{v}}^{-1/2}\left( \sqrt{P_I}\mathbf{H}^I\mathbf{x}^I + \mathbf{n} \right)
\end{eqnarray}
and it is simply treated to follow $\mathcal{CN}{\left(0, \mathbf{I}_{N_r}\right)}$. Consequently, \eqref{eq:IWptp} is equivalent to a point-to-point channel with multiple antennas. Then, the transmit symbol can be detected with a variety of well-known methods such as zero-forcing (ZF) or minimum mean square error (MMSE) equalizers, and maximum likelihood (ML) detector. Under no \emph{a priori} information, the ML detector provides the optimal log-likelihood ratio (LLR) \cite{Hochwald03} for the $n$th bit of the $m$th stream as
\begin{align}
L_{m,n}^{(\textrm{ext})} &= L_{m,n}^{(\textrm{A})} \nonumber \\
&= \ln\frac{P\bracket{b_{m,n} = +1 | \tilde{\mathbf{y}}}}{P\bracket{b_{m,n} = -1 | \tilde{\mathbf{y}}}} \nonumber \\
&= \ln\frac
{\sum_{\substack{\mathbf{b}^D \in \mathbb{B}_{m,n}^{+1}}}P\left(\tilde{\mathbf{y}} | \mathbf{b}^D \right)}
{\sum_{\substack{\mathbf{b}^D \in \mathbb{B}_{m,n}^{-1}}}P\left(\tilde{\mathbf{y}} | \mathbf{b}^D \right)} \nonumber \\
&=\ln\sum_{\substack{\mathbf{x}^D \in \mathbb{X}_{m,n}^{+1}}} \exp\bracket{\mathfrak{H}_{\mathbf{x}}} - \ln\sum_{\substack{\mathbf{x}^D \in \mathbb{X}_{m,n}^{-1}}}\exp\bracket{\mathfrak{H}_{\mathbf{x}}}  \label{eq:IWLLR} \\
&\stackrel{(a)}{\approx} \max_{\mathbf{x}^D \in \mathbb{X}_{m,n}^{+1}}\mathfrak{H}_{\mathbf{x}} - \max_{\mathbf{x}^D \in \mathbb{X}_{m,n}^{-1}}\mathfrak{H}_{\mathbf{x}} \label{eq:IWLLR_approx}
\end{align}
where $L_{m,n}^{(\textrm{ext})}$ and $L_{m,n}^{(\textrm{A})}$ represent the extrinsic and \emph{a posteriori} LLRs, respectively. $\mathbf{b}^{k}$ is the bit vector consisting of all bits corresponding to the symbol element of $\mathbf{x}^{k}$ where $k = D$. This bit vector is generated by aligning all the corresponding bits in a column. Since no \emph{a priori} information is given, the extrinsic LLR is the same as \emph{a posteriori} LLRs. The Euclidean distance for IW is defined as
\begin{align}
\mathfrak{H}_{\mathbf{x}} = -\frac{1}{\sigma_{\tilde{v}}^2} \norm{ \tilde{\mathbf{y}} - \sqrt{P_D}\tilde{\mathbf{H}}^D \mathbf{x}^D }^2
\end{align}
where the variance $\sigma_{\tilde{v}}^2$ is $1$. $\mathbb{B}_{m,n}^{(b)}$ is the set of bits belonging to a transmit symbol vector whose $n$th bit of the $m$th stream is $b$. Similarly, $\mathbb{X}_{m,n}^{(b)}$ indicates the set of transmit symbol vectors where the condition, $b_{m,n} = b$, is satisfied. Both sets are defined as
\begin{align}
\mathbb{B}_{m,n}^{b} &= \{ \mathbf{b}^D | b_{m,n} = b \} \\
\mathbb{X}_{m,n}^{b} &= \{ \mathbf{x}^D | b_{m,n} = b \}.
\end{align}
Beside, $(a)$ is derived with the max-log approximation which states $\ln\sum_i\exp a_i \approx \max_i a_i$. This approximation is used to alleviate the computational burden that is caused by calculating the sum of exponential functions from all constellation points \cite{Hochwald03}. Then, the LLR in \eqref{eq:IWLLR} or the approximated LLR in \eqref{eq:IWLLR_approx} is fed into a decoder after it is de-interleaved.

If an interference signal would be indeed Gaussian distributed rather than modulated on the discrete constellation, the IW could result in the optimal performance. However, in practice, the transmit signal is not Gaussian distributed but modulated on the discrete constellation. This fact becomes the reason to degrade the performance.

%% file: IADet.tex

This subsection reviews the other interference mitigation algorithm to exploit the discrete nature of interference symbols. In \cite{Ghaffar12}, the interference-aware detection (IA-Detection) algorithm is proposed to jointly detect both desired and interference signals. Contrary to the conventional joint detection, the IA-Detection is not required to generate the soft information of both signals but only of the desired signal \cite{Jungwon11}. The desired LLR is directly derived from \eqref{eq:DIeq} and is given by
\begin{align}
L_{m,n}^{(\textrm{ext})} &= L_{m,n}^{(\textrm{A})} \nonumber \\
&= \ln\frac{P\bracket{b_{m,n} = +1 | \tilde{\mathbf{y}}}}{P\bracket{b_{m,n} = -1 | \tilde{\mathbf{y}}}} \nonumber \\
&= \ln\frac
{\sum_{\mathbf{b}^I}\sum_{\substack{\mathbf{b}^D \in \mathbb{B}_{i,j}^{+1}}}P\left(\mathbf{y} | \mathbf{b}^D \mathbf{b}^I \right)}
{\sum_{\mathbf{b}^I}\sum_{\substack{\mathbf{b}^D \in \mathbb{B}_{i,j}^{-1}}}P\left(\mathbf{y} | \mathbf{b}^D \mathbf{b}^I \right)} \nonumber \\
&=\ln\sum_{\mathbf{x}^I}\sum_{\substack{\mathbf{x}^D \in \mathbb{X}_{i,j}^{+1}}} \exp\bracket{\mathfrak{D}_{\mathbf{x}}} - \ln\sum_{\mathbf{x}^I}\sum_{\substack{\mathbf{x}^D \in \mathbb{X}_{i,j}^{-1}}}\exp\bracket{\mathfrak{D}_{\mathbf{x}}} \nonumber \\
&\stackrel{(b)}{\approx} \max_{\substack{\mathbf{x}^D \in \mathbb{X}_{i,j}^{+1}}, \mathbf{x}^I}\mathfrak{D}_{\mathbf{x}} - \max_{\substack{\mathbf{x}^D \in \mathbb{X}_{i,j}^{-1}}, \mathbf{x}^I}\mathfrak{D}_{\mathbf{x}} \label{eq:IADet}
\end{align}
where the Euclidean distance for IA-Detection is defined as
\begin{align}
\mathfrak{D}_{\mathbf{x}} = -\frac{1}{\sigma_n^2} \norm{ \mathbf{y} - \sqrt{P_D}\mathbf{H}^D \mathbf{x}^D - \sqrt{P_I}\mathbf{H}^I \mathbf{x}^I}^2.
\end{align}
This Euclidean distance is calculated on top of modulation information of both desired and interference signals. Similar to the IW, the max-log approximation can be applied to reduce computational complexity in $(b)$. The extrinsic LLR is re-permuted at the de-interleaver and is fed into a decoder. Suppose that both signals having no point-to-point codes are used. Under no \emph{a priori} information, the performance of the IA-Detection algorithm could be maximized with interference modulation information. However, the advent of point-to-point codes for each signal leaves a margin to improve the interference mitigation algorithm.

%% file: IADec.tex

This section introduces two proposed interference-aware decoding algorithms insisting on decoding the interference signal, although each MS is only interested in the desired signal. Since interference channels are different from multiple access channels, the MS does not need to consider how accurately interference signals should be decoded. However, regardless of its accuracy, this paper claims that decoding interference assists to decode the desired signal.

The first proposed algorithm is the interference-aware successive decoding (IASD) that decodes both desired and interference signals successively. Also, another proposed one is the interference-aware parallel decoding (IAPD) that decodes both signals in parallel. The IASD has an advantage to reduce the hardware complexity since only a single decoder is needed to design the receiver. On the other hand, the IAPD can decrease the latency caused by the iteration between a detector and decoders since both signals can be decoded concurrently.

\subsection{Interference-Aware Successive Decoding} \label{sec:IASD}
\input{IASD.tex}

\subsection{Interference-Aware Parallel Decoding} \label{sec:IAPD}
\input{IAPD.tex}

%% file: IASD.tex

Although the IA-Detection presented in \secref{sec:IADet} is well designed with the knowledge of interference modulation, it does not fully exploit all the information of interference. Instead, this algorithm only exploits the property that an interference symbol is one of the modulation symbols discretely arranged on the constellation rather than the symbol following a Gaussian distribution. This property works on the principle that each interference symbol is independent of other symbols. However, interference symbols are originated from the bit sequence coded at the encoder so that they are in fact correlated in a bit-level. Therefore, the joint detection with the knowledge of interference modulation is not enough to alleviate the impact of interference. Instead, the joint decoding with the knowledge of interference coding schemes is required along with the joint detection.

The straightforward method implementing the joint decoding algorithm is to prepare both a series of candidate signal vectors for all message sets and a series of all receive signal vectors, and then to calculate the sum of Euclidean distances between two sets of signal vectors. Lastly, the sum of Euclidean distances needs to be calculated to select the best feasible message set. However, this approach causes extremely computational burden and is beyond the practical range of complexity.

Alternatively, this paper proposes the IASD which decodes the desired signal first and then decodes the interference signal with the updated \emph{a priori} information of the desired signal. Finally, the desired signal is decoded again with the updated \emph{a priori} information of both signals. The proposed algorithm is clearly suboptimal because it is on the way to elaborate the marginal PDF of both desired and interference signals, respectively, rather than the joint PDF of both signals.

\begin{figure*}[htb]
\centering
\includegraphics[width=5.3in]{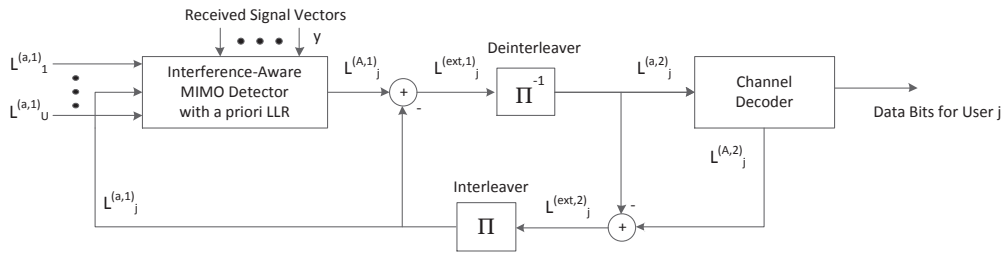}
\caption{A block diagram of soft decoding receiver for user $j$ for interference-aware successive decoding receiver}
\label{fig:receiver}
\end{figure*}

\figref{fig:receiver} shows the block diagram of an IASD architecture, consisting of the interference-aware detector supported by \emph{a priori} information and the channel decoder. The interference-aware detector is extended to incorporate \emph{a priori} LLRs of both desired and interference signals, which are given by
\begin{eqnarray}
L_{m,n}^{(a,1,D)} &=& \ln\frac{P\left(b_{m,n}^D = +1\right)}{P\left(b_{m,n}^D = -1\right)}, \label{eq:ap_D}\\
L_{m,n}^{(a,1,I)} &=& \ln\frac{P\left(b_{m,n}^I = +1\right)}{P\left(b_{m,n}^I = -1\right)} \label{eq:ap_I}
\end{eqnarray}
where the superscript $1$ indicates that these LLRs are associated with the detector, not the decoder. As shown in \figref{fig:receiver}, \emph{a priori} information associated with the decoder is expressed with the superscript $2$. Thus, $L_{m,n}^{(a,1,D)}$ and $L_{m,n}^{(a,1,I)}$ denote \emph{a priori} LLR of the desired and interference signal, serving for the detector corresponding to the $n$th bit of the $m$th stream, respectively. Given the received signal, the detector generates \emph{a posteriori} LLRs that is defined as
\begin{eqnarray}
L_{m,n}^{(A,1,k)} = \ln\frac{P\left(b_{m,n}^k = +1 | \mathbf{y}\right)}{P\left(b_{m,n}^k = -1  | \mathbf{y}\right)}
\end{eqnarray}
where the superscript $k$ could be $D$ or $I$ depending on the detecting order. The extrinsic LLR is given by eliminating the dependency of \emph{a priori} LLR from \emph{a posteriori} LLR. In detail, the extrinsic LLR for the desired signal can be expanded into
\begin{align}
L_{m,n}^{(\textrm{ext},1,D)} &= L_{m,n}^{(\textrm{A},1,D)} - L_{m,n}^{(\textrm{a},1,D)} \nonumber \\
&\hspace{-1.0cm}= \ln\frac
{\sum_{\mathbf{b}^I}\sum_{\substack{\mathbf{b}^D \in \mathbb{B}_{m,n}^{+1}}}P\left(\mathbf{y} | \mathbf{b}^D \mathbf{b}^I \right) P\left(\mathbf{b}^D_{[m,n]}, \mathbf{b}^I\right)}
{\sum_{\mathbf{b}^I}\sum_{\substack{\mathbf{b}^D \in \mathbb{B}_{m,n}^{-1}}}P\left(\mathbf{y} | \mathbf{b}^D \mathbf{b}^I \right) P\left(\mathbf{b}^D_{[m,n]}, \mathbf{b}^I\right) } \nonumber\\
&\hspace{-1.0cm}\stackrel{(c)}{=}\ln\sum_{\mathbf{x}^I}\sum_{\substack{\mathbf{x}^D \in \mathbb{X}_{m,n}^{+1}}} \exp\left(\mathfrak{D}_{\mathbf{x}}  + \mathfrak{L}_{m,n} \right) \nonumber \\
&\hspace{-1.0cm}- \ln\sum_{\mathbf{x}^I}\sum_{\substack{\mathbf{x}^D \in \mathbb{X}_{m,n}^{-1}}}\exp\left(\mathfrak{D}_{\mathbf{x}}  + \mathfrak{L}_{m,n} \right) \label{eq:IASD1} \\
&\hspace{-1.0cm}\approx \max_{\substack{\mathbf{x}^D \in \mathbb{X}_{m,n}^{+1} \mathbf{x}^I}}
\left(\mathfrak{D}_{\mathbf{x}}  + \mathfrak{L}_{m,n}\right) \nonumber \\
&\hspace{-1.0cm}- \max_{\substack{\mathbf{x}^D \in \mathbb{X}_{m,n}^{-1}, \mathbf{x}^I}}
\left(\mathfrak{D}_{\mathbf{x}}  + \mathfrak{L}_{m,n}\right)  \label{eq:IASD}
\end{align}
where $\mathbf{b}_{[m,n]}^D$ is the subset of $\mathbf{b}^D$ excluding the $n$th bit of the $m$th stream. The sum of bit vectors used in \eqref{eq:IASD1} and \eqref{eq:IASD} are given by
\begin{align}
\mathfrak{L}_{m,n} = \frac{1}{2} \mathbf{b}^{I\dagger} \mathbf{L}^{(a,1,I)} + \frac{1}{2} \mathbf{b}_{[m,n]}^{D\dagger} \mathbf{L}^{(a,1,D)}_{[m,n]}
\end{align}
where $\mathbf{L}^{(a,1,k)}$ is the vector of \emph{a priori} LLRs corresponding to $\mathbf{b}^{k}$ with the same superscript. In $(c)$, the joint bit probability of $\mathbf{b}^D$ and $\mathbf{b}^I$ is developed by using the fact that both signals are independent and the bit probability can be converted from \eqref{eq:ap_D} and \eqref{eq:ap_I} to
\begin{eqnarray}
P(b_{m,n} = b) = \frac{\exp\left(b \frac{L_{m,n}}{2}\right)}{\exp\left(\frac{L_{m,n}}{2}\right) + \exp\left(-\frac{L_{m,n}}{2}\right)}
\end{eqnarray}
where $b \in \{\pm 1\}$. Once the extrinsic LLRs are obtained, they are de-interleaved and become robust to the channel variation. Then, the re-permuted extrinsic LLRs are turned into soft inputs for the channel decoder as
\begin{eqnarray}
L_{m,n}^{(a,2,k)} = \Pi^{-1}\left(L_{m,n}^{(\textrm{ext},1,k)}\right),
\end{eqnarray}
which are \emph{a priori} LLRs for the decoder. Similarly, the channel decoder also produces its own \emph{a posteriori} LLRs and the extrinsic LLRs are accordingly calculated as
\begin{eqnarray}
L_{m,n}^{(\textrm{ext},2,k)} = L_{m,n}^{(A,2,k)} - L_{m,n}^{(a,2,k)}.
\end{eqnarray}
Using the feedback path to the detector, this extrinsic information is interleaved and served for \emph{a priori} information at the detector as
\begin{eqnarray}
L_{m,n}^{(a,1,k)} = \Pi\left(L_{m,n}^{(\textrm{ext},2,k)}\right).
\end{eqnarray}
The iterative procedure continues until the stopping criterion is satisfied. Due to the complexity, the IASD restricts the number of iteration to decode the desired signal twice and the interference signal once. Since this procedure is repeated successively, only one decoder is enough to run the IASD.

%% file: IAPD.tex

\begin{figure*}[htb]
\centering
\includegraphics[width=5.3in]{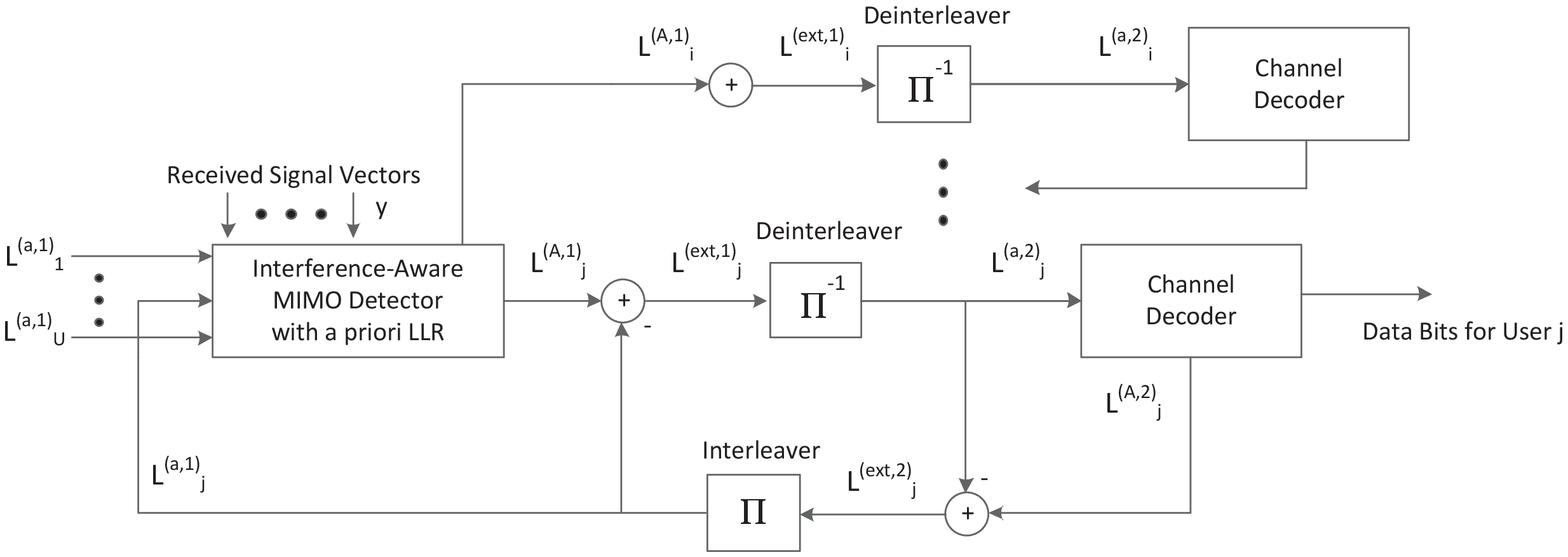}
\caption{A block diagram of soft decoding receiver for user $j$ for interference-aware parallel decoding receiver}
\label{fig:receiver_IAPD}
\end{figure*}

Under no constraints of complexity, multiple decoders can be simultaneously used at the receiver architecture. The proposed IASD can be extended to use multiple decoders that enable to decode all streams from the detector in parallel, called the interference-aware parallel decoding (IAPD) compared to the IASD. The proposed IAPD should change the detector from an interference-aware detector to a joint detector that generates the soft outputs of all streams at once. \figref{fig:receiver_IAPD} depicts the block diagram of an IAPD receiver architecture using multiple decoders with multiple feedback paths. The IAPD operates in the same way as the IASD except the fact that the extrinsic LLRs of all streams are updated simultaneously. Thus, the receiver is required to tolerate the hardware complexity caused by multiple decoders.

%% file: complexity.tex
\input{EXIT.tex}

%% file: EXIT.tex
Using the extrinsic information transfer (EXIT) chart introduced in \cite{tenBrink2001}, this section demonstrates the superiority of the proposed interference-aware decoding algorithm compared to non-decoding algorithms. The EXIT chart is a versatile tool to analyze the receiver with the IDD structure over BICM \cite{tenBrink2000,tenBrink2001,Hagenauer04theexit}. The IDD receiver is designed with both feedforward and feedback paths between a detector and a decoder. In order to research the error events generated at this IDD receiver, its information flow needs to be tracked accordingly. For this purpose, the pairwise error probability (PEP) \cite{Caire98} or the belief propagation (BP) message-passing algorithm based on factor graph \cite{Kschischang01} can be used to derive the probability of errors that the set of messages or codewords are incorrectly decoded. However, these approaches are inevitably complex because both the feedforward path from the detector to the decoder and the feedback path from the decoder to the detector should be simultaneously investigated for studying the probability of errors. Moreover, the PEP is easy to be loose due to the fact that the PEP only considers the union bound of error events in nature.

Alternatively, the EXIT chart enables the investigation on the IDD receiver to be divided into two parts for the detector and the decoder, respectively. Both blocks should be concurrently analyzed to reveal the relation between soft inputs and soft outputs based on mutual information. The soft input and output correspond to \emph{a priori} LLRs and extrinsic LLRs for each block, respectively. Under the assumption that (de-)interleaved extrinsic LLRs are i.i.d., the EXIT chart describes the flow of extrinsic information given \emph{a priori} information. Besides, the trajectory on the EXIT chart can visualize the exchange of extrinsic information between the detector and the decoder.

To draw the EXIT chart, the LLR sequence needs to be generated to follow its distribution. Suppose that the channel with a bipolar input bit $x \in \{ \pm 1\}$ is given as
\begin{eqnarray}
y = h x + n
\end{eqnarray}
where $h$ is the complex channel gain and the noise $n$ follows a complex Gaussian distribution with mean zero and variance $\sigma_n^2$. Given the channel, \emph{a posteriori} LLR is simply calculated as
\begin{align}
L^{(A)} &= \ln\frac{P(x = +1|y)}{P(x = -1|y)} \\
&= \ln\frac{P(x = +1)}{P(x = -1)} + \frac{h^*y + y^*h}{\sigma_n^2} \\
&= L^{(a)} + \frac{h^*y + y^*h}{\sigma_n^2} \label{eq:exit_channel}
\end{align}
which is developed with \emph{a priori} LLR and the extrinsic LLR. In \eqref{eq:exit_channel}, the extrinsic LLR is expressed as the the difference between $L^{(A)}$ and $L^{(a)}$, and can be re-defined with the input bit $x$ and the channel state gain $\mu$ as
\begin{align}
L^{(\textrm{ext})} &= \frac{2\abs{h}^2}{\sigma_n^2}x + \tilde{n} \\
&= \mu x + \tilde{n}. \label{eq:exit_ch_state}
\end{align}
where the variance of the complex Gaussian noise $\tilde{n}$ is given by
\begin{eqnarray}
\sigma_{\tilde{n}}^2 = \frac{4\abs{h}^2}{\sigma_n^2} = 2\mu.
\end{eqnarray}
In \cite{Hagenauer96}, it is shown that the variance $\sigma_{\tilde{n}}^2$ satisfies the condition that $\sigma_{\tilde{n}}^2 = 2\mu$. The extrinsic sequence of the detector is scrambled with the de-interleaver as $\tilde{x} = \Pi^{-1}\left( x \right)$ and becomes \emph{a priori} sequence serving as the input of the decoder in a feedforward path. When the feedback path is concerned, the detector and the decoder switch their roles to that the scrambled extrinsic sequence of the decoder becomes \emph{a priori} sequence of the detector.

In the EXIT chart analysis \cite{Hagenauer04theexit}, \emph{a priori} information can be modeled by a Gaussian distributed sequence:
\begin{eqnarray}
L^{(a)} = \mu \tilde{x} + \tilde{n} \label{eq:La_generating}
\end{eqnarray}
where the binary random variable $\tilde{X}$ denotes the input bits with the realization $\tilde{x} \in \{ \pm 1\}$. Then, the conditional PDF of \emph{a priori} LLR given $\tilde{x}$ can be expressed with the variance $\sigma_{\tilde{n}}^2$ as
\begin{align}
P_{L^{(a)}}\left(L | \tilde{X} = \tilde{x}\right) &\nonumber \\
&\hspace{-1cm}= \frac{1}{\sqrt{2 \pi \sigma_{\tilde{n}}^2}} \exp{\left(-\frac{1}{2\sigma_{\tilde{n}}^2}\abs{L - \frac{\sigma_{\tilde{n}}^2}{2}\tilde{x}}^2\right)}. \label{eq:condPDF}
\end{align}
Using this conditional PDF, the amount of \emph{a priori} knowledge can be measured with mutual information between the scrambled input bit $\tilde{x}$ and  \emph{a priori} LLR, and is expressed as $I_{L^{(a)}} = I\left( \tilde{X}; L^{(a)}\right)$. As shown in \eqref{eq:condPDF}, this mutual information is a function of the noise variance $\sigma_{\tilde{n}}$ such that it is calculated as \cite{tenBrink2000,Hagenauer04theexit}
\begin{align}
I_{L^{(a)}}\left(\sigma_{\tilde{n}}\right) & \nonumber \\
&\hspace{-1cm}= 1 - \int_{-\infty}^{\infty}P_{L^{(a)}}\left(L | \tilde{X} = +1\right)\log_2{\left(1 + e^{-L}\right)} dL \\
&\hspace{-1cm}= 1 - E\left[ \log_2{\left(1 + e^{-L}\right)}\right]. \label{eq:MI_1}
\end{align}
Moreover, by using the change of variables, $I_{L^{(a)}}\left(\sigma_{\tilde{n}}\right)$ can be rewritten as
\begin{align}
I_{L^{(a)}}\left(\sigma_{\tilde{n}}\right) & \nonumber \\
&\hspace{-1cm}= 1 - \int_{-\infty}^{\infty} \frac{1}{\sqrt{2 \pi}} \exp{\left(-\frac{ z^2 }{ 2 }\right)} \nonumber \\
&\hspace{+1cm}\log_2{\left(1 + \exp{\left(-\left[\sigma_{\tilde{n}}z + \frac{\sigma_{\tilde{n}}^2}{2}\right]\right)}\right)} dt \\
&\hspace{-1cm}= 1 - E\left[ \log_2{\left(1 + \exp{\left(-\left[\sigma_{\tilde{n}}z + \frac{\sigma_{\tilde{n}}^2}{2}\right]\right)}\right)} \right] \label{eq:MI_2}
\end{align}
where $z$ is a normal Gaussian random variable with $\mathcal{N}(0,1)$. Therefore, the mutual information is expressed as the expectation of the logarithm of the LLR sequence as in \eqref{eq:MI_1} or is specified with the variance $\sigma_{\tilde{n}}$ as in \eqref{eq:MI_2}. Using the function $J(\sigma)$ defined in \cite{tenBrink2000} as
\begin{align}
J(\sigma) &\triangleq I_{L^{(a)}}\left(\sigma_{\tilde{n}} = \sigma\right) \\
\sigma_{\tilde{n}} &= J^{-1}\left(I_{L^{(a)}}\right),
\end{align}
the variance $\sigma_{\tilde{n}}$ can be reversely calculated from the mutual information of \emph{a priori} information, and consequently \emph{a priori} LLR sequence can be generated from \eqref{eq:La_generating}. This sequence serves as the input of the decoder or the detector, and correspondingly generates the output extrinsic LLR sequence mapped to the mutual information $I_{L^{(\textrm{ext})}} = I\left( \tilde{X}; L^{(\textrm{ext})}\right)$.


\begin{figure}[htb]
\centering
\includegraphics[width=3.2in]{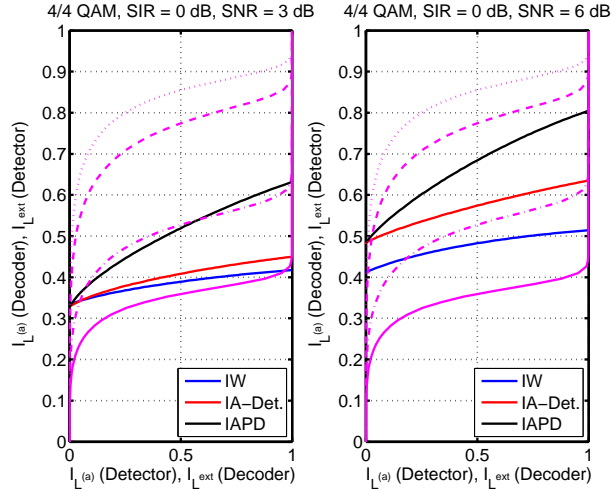}
\caption{EXIT charts mapping the proposed IAPD along with interference-whitening and interference-aware detectors. The extrinsic information curves for the decoder corresponds to the rates, $0.83, 0.75, 0.50$ and $0.33$, in the order from the top. Both desired and interference signals used $4$ QAM at SIR $= 0$ dB.}
\label{fig:EXIT1}
\end{figure}

\begin{figure}[htb]
\centering
\includegraphics[width=3.2in]{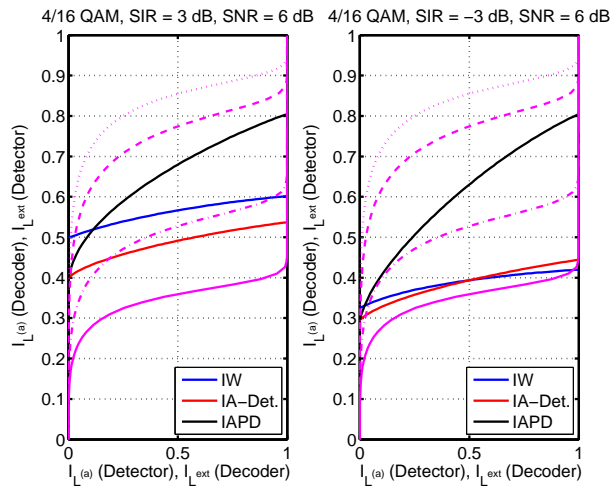}
\caption{EXIT charts mapping the proposed IAPD along with interference-whitening and interference-aware detectors. The extrinsic information curves for the decoder corresponds to the rates, $0.83, 0.75, 0.50$ and $0.33$, in the order from the top. The desired signal used $4$ QAM. On the other hand, the interference signals used $16$ QAM at SNR $= 6$ dB.}
\label{fig:EXIT2}
\end{figure}

\figref{fig:EXIT1} and \figref{fig:EXIT2} show the EXIT charts of two interference mitigation algorithms described reviewed in \secref{sec:algorithm} with the proposed IAPD in \secref{sec:IAPD}. Both desired and interference signals are modulated on the $4$ QAM constellation in \figref{fig:EXIT1}, and the interference modulation is changed to $16$ QAM in \figref{fig:EXIT2}. In each figure, the detector extrinsic curves are combined with the inverted decoder extrinsic curves with $4$ code rates, $0.33, 0.50, 0.75$ and $0.83$, from the bottom to the top in a magenta color. Since the proposed IAPD simultaneously updates both \emph{a priori} information of desired and interference signals, the detector extrinsic curve corresponding to the IAPD denotes the mutual information of either the desired or the interference signal.

First, in \figref{fig:EXIT1}, the curves are plotted with the SIR $0$ dB and two SNR levels: $3$ and $6$ dB. It is observed that the proposed IAPD outperforms both non-decoding algorithms, IW and IA-Detection. For example, at SNR $6$ dB, the IW cannot support the code rate $0.50$ because the detector extrinsic curve and the inverted decoder extrinsic curve are overlapped each other. Even for the IA-Detection, the trajectory of the extrinsic information requires more than $4$ iterations to reach at a reliable status where the mutual information becomes $1$. However, the proposed IAPD only needs $2$ iterations to be fully reliable so as to show its superiority.

\figref{fig:EXIT2} depicts the EXIT chart at SNR $6$ dB, but with different SIR levels, $3$ and $-3$ dB. When the SIR level decreases, it is shown that the extrinsic information gap between the IAPD and non-decoding algorithms increases. On the contrary, when the SIR level increases, the gap is shown to decrease. The strong interference signal is rather advantageous to the proposed IAPD because it is easy to be decoded. However, IW and IA-Detection are not easy to decode the desired signal when the SIR level is high. Besides, it is observed that at SIR $3$ dB, IW initially produces more reliable mutual information than IAPD. In \eqref{eq:IASD}, the max-log approximation is used to alleviate the computational burden. Since the proposed IASD and IAPD detect more streams than IW, the approximation error would cause to reverse the performance at the low mutual information region.

%% file: results.tex

This section evaluates the proposed interference-aware decoding algorithm in terms of modulation orders, coding rates, SIRs and SNRs. Besides, the analytical results in \secref{sec:analysis} are compared with simulation results generated by MATLAB. The number of BSs and MSs is set to be $2$. The number of transmit, receive antennas and spatial streams is $N_t = N_r = N_s = 2$, respectively. The elements of channel matrices follow a Rayleigh fading distribution. This section assumes that the MS can estimate the channel matrix perfectly. The transmitted symbols are modulated on the QAM constellation with Gray mapping. A single packet consists of $400$ data bits that are encoded with turbo codes generated with the polynomials $(7,5)$. Then, the coded packet, having $400 / R_c$ where $R_c$ is the code rate, is transmitted over $10$ subcarriers with $2$ codewords. Each of systematic and parity bits in a single packet are permuted with their own random interleavers, respectively. The performance of each algorithm is measured with packet error rates (PERs) over $10000$ packets. For comparison with non-iterative algorithms, both IW and IA-Detection are simulated with $8$ inner iterations at the turbo decoder, while IASD and IAPD are simulated with $4$ inner iterations at the decoder and $2$ outer iterations per codeword with the detector. In detail, the desired signal is repeatedly decoded twice and the interference signal is decoded only once in the IASD.

\begin{figure}[htb]
\centering
\includegraphics[width=3.2in]{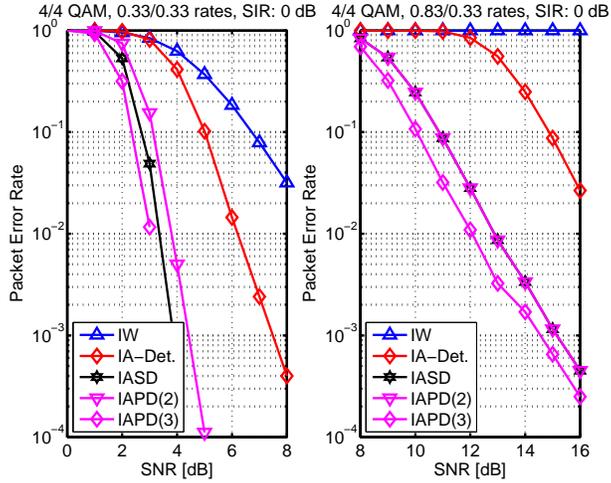}
\caption{The PERs are plotted with different desired code rates, $0.33$ and $0.83$, at SIR $0$ dB.}
\label{fig:4QAM_d0.83}
\end{figure}

\begin{figure}[htb]
\centering
\includegraphics[width=3.2in]{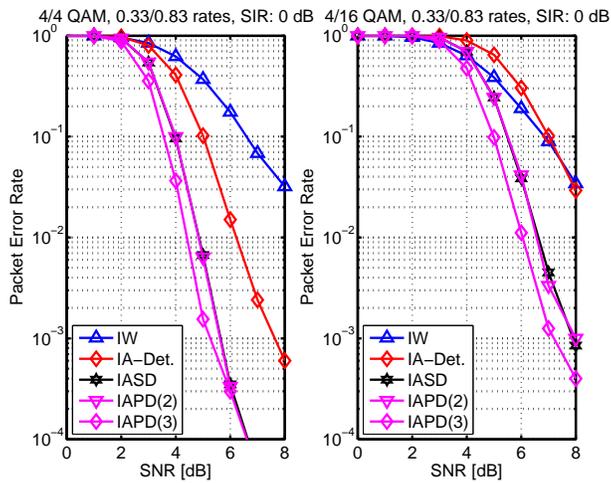}
\caption{The PERs are plotted with different interference modulation orders, $4$ and $16$ QAM, at SIR $0$ dB.}
\label{fig:16QAM_i0.83}
\end{figure}

\figref{fig:4QAM_d0.83} and \figref{fig:16QAM_i0.83} compare the proposed IASD and IAPD with IW and IA-Detection in terms of coding rates and modulation orders. The legend `IAPD$(n)$' means that $n$ outer iterations are used where $n = 2$ or $3$. All curves are plotted with SIR $0$ dB, i.e., $P_D = P_I$. For IW, it is shown that the PER only depends on the desired code rate. Since the IW treats interference as Gaussian noise, its modulation order and code rates should be independent of the PER. Accordingly, the PER of IA-Detection is only changed when the interference modulation order is changed from $4$ to $16$ QAM. It is observed to be independent of the interference code rate. On the other hand, the proposed IASD and IAPD are designed to decode interference signals and to use the decoding results as the updated \emph{a priori} information. Hence, the PER of IASD and IAPD depends on both interference modulation order and code rates.

As the desired code rate increases from $0.33$ to $0.83$, \figref{fig:4QAM_d0.83} depicts that the gain of IASD and IAPD over IW and IA-Detection also increases. This implies that the reduced redundancy at the code rate $0.83$ can be compensated from the updated \emph{a priori} information at the proposed IASD and IAPD. As explained in \secref{sec:analysis}, the approximation error used for deriving \eqref{eq:IADet} and \eqref{eq:IASD} could cause to reverse the performance at a high BLER region. It is detected that this phenomenon occurs at $16$ QAM with code rate $0.83$ for interference signals. However, at most of SNR regions, the superiority of the proposed IASD and IAPD is clearly verified without any errors.

\begin{figure}[htb]
\centering
\includegraphics[width=3.2in]{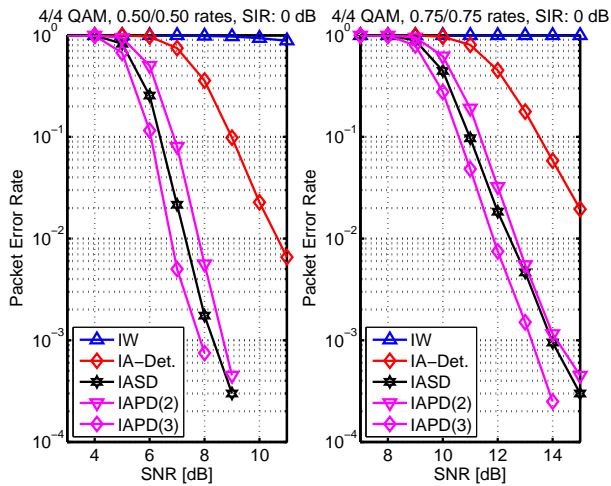}
\caption{The PERs are plotted with different code rates, $0.50$ and $0.75$, at SIR $0$ dB.}
\label{fig:4QAM_di0.50and0.75}
\end{figure}

\figref{fig:4QAM_di0.50and0.75} plots the PER using high code rates $0.50$ and $0.75$ for both desired and interference signals. In particular, these PERs can be compared with the EXIT chart for SNR $3$ and $6$ dB in \figref{fig:EXIT1}. The EXIT chart is derived on the theoretical assumption that the \emph{a priori} sequence is infinite and follows a Gaussian distribution. On the other hand, simulation results are achieved by the finite \emph{a priori} sequence and the discrete modulation order. Nonetheless, \figref{fig:4QAM_di0.50and0.75} is well matched with the EXIT chart in \figref{fig:EXIT1}. For instance, the proposed IAPD cannot support the code rate $0.75$, which is also verified in \figref{fig:4QAM_di0.50and0.75}. At SNR $6$ dB, the IAPD trajectory with $2$ iterations on the EXIT chart cannot reach at a reliable status. However, with $3$ iterations, it becomes reliable. The PER in \figref{fig:4QAM_di0.50and0.75} is consistent with this interpretation.

\begin{figure}[htb]
\centering
\includegraphics[width=3.2in]{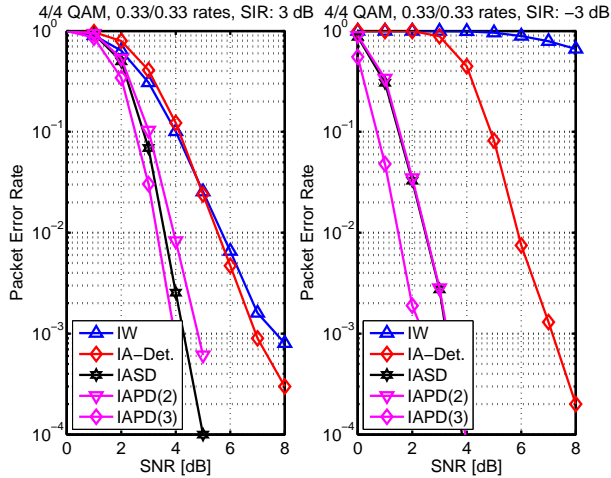}
\caption{The PERs are plotted with different SIRs $\pm 3$ dB.}
\label{fig:4QAM_SIR3_-3}
\end{figure}

\figref{fig:4QAM_SIR3_-3} demonstrates the PER of different SIR levels, $\pm 3$ dB. It is shown that the gain of IASD and IAPD over IW and IA-Detection increases as the SIR level decreases. Intuitively, strong interference channels are favorable for IASD and IAPD due to ease of interference decodability. This implies that the proposed IASD and IAPD are robust against the strength of interference channels, while IW and IA-Detection are not.

\begin{figure}[htb]
\centering
\includegraphics[width=3in]{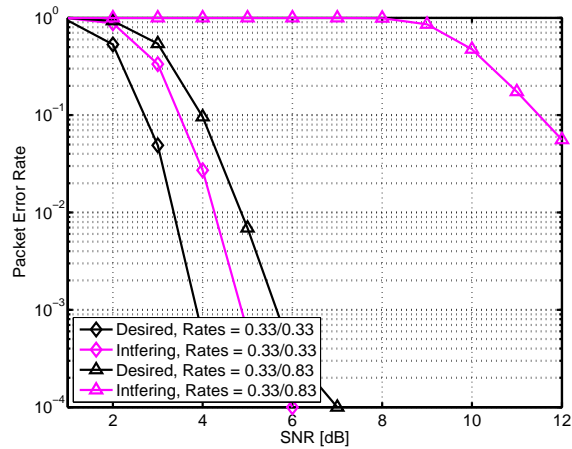}
\caption{The PERs of the desired and interference signal for IASD are separately plotted.}
\label{fig:derired_and_interference}
\end{figure}

\figref{fig:derired_and_interference} compares the PER of both desired and interference signals when the proposed IASD is used. As explained for IASD, the desired signal is decoded twice while the interference signal is decoded once. Although the same code rates are used for both signals, the PER of the desired signal should outperform the PER of the interference signal. As the interference code rate increases, the PER difference between both signals becomes larger accordingly. The proposed IASD is not intended to achieve the performance of interference signals but only takes care of the performance of desired signals. This issue essentially distinguishes interference channels from multiple access channels where a receiver is interested in decoding both desired and interference signals.

\begin{figure}[htb]
\centering
\includegraphics[width=3in]{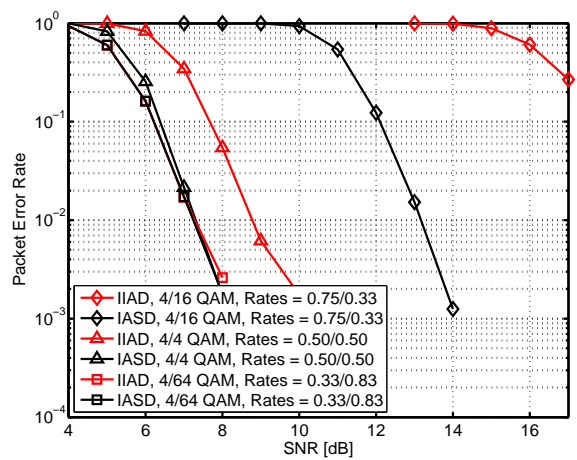}
\caption{The PERs of IASD and IIAD are compared where IIAD means IASD without decoding interference signals.}
\label{fig:signal_iteration}
\end{figure}

\figref{fig:signal_iteration} plots the PER of IASD as well as the PER of iterative interference-aware detecting (IIAD) that is repeatedly decoding the desired signal and updating \emph{a priori} information. Similar to IASD, IIAD also employs interference-aware detection in order to generate the extrinsic LLRs of the desired signal. Hence, the difference between IASD and IIAD arises from the gain of interference decoding. \figref{fig:signal_iteration} also indicates that interference decoding does not hurt the proposed IASD by showing that IIAD cannot outperform IASD. For example, although interference signals are modulated with $64$ QAM and code rate $0.83$, interference decoding at least does not hurt IASD even with inaccurate \emph{a priori} information of interference signals. Instead, when interference signals are successfully decoded, the IASD gain can be much bigger.

%% file: conclusion.tex

This paper investigated the problem of mitigating interference in multi-input multi-output interference channels with point-to-point codes and BICM. The interference-aware successive decoding algorithm was proposed as a practical way of implementing joint decoding of the desired and interference signals. Using the IASD and IAPD algorithms, it was shown that interference should not be simply treated as noise and also the interference detecting was not enough to achieve the performance of high throughput. Instead, it was shown with both the EXIT charts and simulation results that the interference decoding was critical to improve the performance. The interference decoding needs the interference information of modulation as well as its coding rate. Thus, the further research to support interference information to MSs is required in the latest standards such as the recent coordinated multi-point transmission (CoMP) standard.